\documentclass[%
  aps,
  pre,
  superscriptaddress, letterpaper,
  longbibliography,
  showpacs,
  showkeys,
  amsfonts, amssymb, amsmath]{revtex4-1}

\usepackage[utf8]{inputenc}

\usepackage[hidelinks]{hyperref}
\usepackage{hypernat}   

\usepackage{microtype}
\usepackage{graphicx}
\usepackage[svgnames]{xcolor}

\usepackage[normalem]{ulem}
\usepackage{siunitx}    
\newcommand*{\TITLE}{%
  Lagrangian acceleration statistics in a turbulent channel flow
}

\hypersetup{%
  pdftitle=\TITLE,%
}

\newcommand*{\Rey}{\mathit{Re}}  
\newcommand*{\Retau}{\Rey_\tau}         

\newcommand*{\yplus}{y^+}

\newcommand*{\mean}[1]{\langle {#1} \rangle}

\newcommand*{\anorm}{{\lvert \vec{a} \rvert}}  

\renewcommand*{\vec}[1]{\mathbf{#1}}

\begin{document}

\begin{abstract}
  Lagrangian acceleration statistics in a fully developed turbulent channel flow
  at $\Retau = \num{1440}$ are investigated, based on
  tracer particle tracking in experiments and direct numerical simulations.
  The evolution with wall distance of the Lagrangian velocity and acceleration
  time scales is analyzed.
  Dependency between acceleration components in the near-wall region is
  described using cross-correlations and joint probability density functions.
  The strong streamwise coherent vortices typical of wall-bounded turbulent
  flows are shown to have a significant impact on the dynamics.
  This results in a strong anisotropy at small scales in the near-wall region that
  remains present in most of the channel.
  Such statistical properties may be used as constraints in building advanced
  Lagrangian stochastic models to predict the dispersion and mixing of chemical
  components for combustion or environmental studies.
\end{abstract}

\title{\TITLE}
\date{\today}



\newcommand*{\LEGI}{%
  Laboratoire des Ecoulements G\'eophysiques et Industriels,
  Universit\'e Grenoble Alpes \& CNRS,
  Domaine Universitaire, CS 40700, F-38058 Grenoble, France
}

\newcommand*{\LMFA}{%
  Laboratoire de M\'ecanique des Fluides et d'Acoustique, UMR 5509,
  Ecole Centrale de Lyon, CNRS, Universit\'e Claude Bernard Lyon 1, INSA Lyon,
  36 av. Guy de Collongue, F-69134 Ecully, France
}

\author{Nickolas Stelzenmuller}
\affiliation{\LEGI}

\author{Juan Ignacio \surname{Polanco}}
\affiliation{\LMFA}

\author{Laure Vignal}
\affiliation{\LEGI}

\author{Ivana Vinkovic}
\affiliation{\LMFA}

\author{Nicolas Mordant}
\email[]{nicolas.mordant@univ-grenoble-alpes.fr}
\affiliation{\LEGI}

\maketitle

\section{Introduction}

The Lagrangian study of fluid particle trajectories is a natural first step in predicting
the transport of components that are passively entrained by the flow such as
chemical/radioactive pollution (passive scalars), aerosols (with small
inertia) or the mixing of components prior to combustion. Indeed, in many cases
the P\'eclet number is very large so that most of the statistical properties of
scalar dispersion are directly related to that of the dispersion of fluid
particles (except of course at the smallest scales at which molecular diffusion
plays an ultimate role that depends on the Schmidt number). Despite
advances in computational power and resources, direct numerical simulations (DNS) are
still out of reach in practical situations.
Thus, an efficient modeling is
required to obtain reliable predictions. Due to the random nature of turbulence,
it is tempting to develop stochastic
models that could be used for simulations with an average or large-scale
knowledge of the flow. A growingly popular method is Large Eddy Simulation (LES)
in which only the largest scales of the turbulence are resolved whereas the small
scales of the turbulent spectra are modeled~\cite{sagaut_large_2006}. Various
classes of models can be used for the unresolved part of the flow, and Lagrangian
stochastic subgrid models can be developed for such simulations as used in
combustion
for instance~\cite{pope_pdf, zamansky_acceleration_2013}.
An efficient model would also be useful to forecast the dispersion of pollution
from localized sources (e.g.\ industrial accident) using coarse grid
meteorological predictions.

In the context of homogeneous and isotropic turbulence (HIT), 1D Lagrangian
stochastic models have been developed as variations of the Langevin equation,
i.e.\ modeling the velocity $v$ of a fluid particle as a Markovian
process~\cite{Pope_pdf2}:
\begin{equation}
  dv=-\frac{v}{T_L}dt + {\left( \frac{2\sigma^2}{T_L} \right)}^{1/2} dW(t)
 \label{eq:langevin}
\end{equation}
with $T_L$ the Lagrangian integral time scale, $W(t)$ a Wiener process and
$\sigma^2$ the velocity variance. Due to the absence of correlations between
velocity components this equation is 1D. It is so strongly
constrained by symmetries and the input from the Kolmogorov 1941
theory that it involves only one parameter $T_L=\frac{2\sigma^2}{C_0\epsilon}$
(with $\epsilon$ the average turbulent energy dissipation rate per unit mass and
$C_0$ a universal constant). This
approach incorporates naturally Taylor's classic
result~\cite{taylor_diffusion_1920} of long term turbulent diffusion of a single
particle. This equation includes neither the dependency on the Reynolds number
nor intermittency. Concerning the former point, this simple framework
(equation~\ref{eq:langevin}) has been
extended by \textcite{sawford_reynolds_1991} to include finite Reynolds number
effects. The model is now a second order stochastic equation that models the
acceleration and no longer the velocity:
\begin{equation}
  da = -\alpha_1 a dt - \alpha_2 \int_0^t a(s) ds dt +
  \sqrt{2\alpha_1\alpha_2\sigma^2} dW(t)\, .  \label{eq:saw91}
\end{equation}
Parameters $\alpha_1$ and $\alpha_2$ are two inverse time scales related to
the Kolmogorov time scale $\tau_\eta=\sqrt{\nu/\epsilon}$ and the integral
Lagrangian time scale $T_L$. The Reynolds number thus appears as the ratio of
the two time scales. At very high Reynolds number, the Kolmogorov 1941 theory
predicts that the ratio $\frac{T_L}{\tau_\eta}=\frac{2Re_\lambda}{C_0\sqrt{15}}$
(with $Re_\lambda$ the usual Taylor-scale Reynolds number).
Sawford~\cite{sawford_reynolds_1991} suggested
an empirical formula estimated from DNS at moderate $Re_\lambda$ given by:
\begin{equation}
    \frac{T_L}{\tau_\eta}=\frac{2Re_\lambda}{C_0\sqrt{15}}(1+7.5C_0^2Re_\lambda^{-1.64}).
    \label{eq:HIT_model}
\end{equation}
This model remains Gaussian at all scales and thus does not include
any intermittency effect.
Moreover, it remains unidimensional with no
interdependency between acceleration (or velocity) components. Real flows as
encountered in nature (atmospheric boundary layer) or in industrial applications (pipes,
mixers, combustion chambers\ldots) can rarely be considered as homogeneous and
isotropic, notably due to
non-zero average shear and wall-confinement.
\textcite{pope_stochastic_2002-1} showed that a homogeneous, anisotropic
Langevin-type stochastic model should include the time scales associated with
the auto- and cross-correlations of the acceleration and velocity components.

The last twenty years have seen the development of experimental techniques and
DNS capabilities that have allowed the direct
simultaneous observation of the acceleration, velocity, and position of fluid
particles, mostly focused on HIT.
They showed that acceleration statistics are strongly non-Gaussian
(intermittent)~\cite{la_porta_fluid_2001}.
Modelling such features requires to further increase the dimensionality of the
stochastic models in the framework of
the non extensive statistical mechanics~\cite{PopeChen,Beck,Reynolds} or to use
a non-Markovian model~\cite{mordant_long_2002}. In both cases, inspired by the
Kolmogorov-Obukhov 1962 theory, dissipation (that appears in the magnitude of
the noise in the stochastic equations) is assumed to be itself a stochastic
variable and fluctuates with a long time scale comparable to $T_L$. Thus, the
stochastic equations involve multiplicative noise that make their developments
much more involved.

Data concerning more realistic flows are
scarce~\cite{choi_lagrangian_2004,chen_acceleration_2010,
walpot_determination_2007,Castello_Clercx,Gerashchenko, taniere_study_2010,
kuerten_lagrangian_2013}.
Complex models of inhomogeneous and anisotropic turbulence are weakly
constrained by symmetries or scaling considerations and thus require significant
experimental or numerical input. 
Del Castello \& Clercx~\cite{Castello_Clercx} studied anisotropic turbulence
affected by rotation which remains quite far from realistic flows. Walpot {\it
et al.} reported some Lagrangian statistics of velocity in the circular pipe
flow and their incorporation in stochastic modeling but no acceleration
data~\cite{walpot_determination_2007}. Gerashenko {\it et al.}
\cite{Gerashchenko} studied the case of inertial (heavy) particles in a boundary
layer but not the case of the Lagrangian tracers. Chen {\it et
al.}~\cite{chen_acceleration_2010} provided Eulerian information on the
acceleration in a turbulent channel flow but did not discuss the Lagrangian
dynamics. Choi et al.~\cite{choi_lagrangian_2004} report a numerical analysis of
the Lagrangian dynamics of acceleration in a turbulent channel flow but their
Reynolds number is relatively low and they discuss neither the coupling between
acceleration components nor the time scales.
This article reports small
scale-resolved Lagrangian experimental
measurements in a statistically stationary, high aspect ratio turbulent channel flow, as well
as DNS results with parameters matching those of the experiment.
Such a flow represents a relatively simple academic framework that incorporates
the basic ingredients of real flows: average shear (anisotropy) and confinement
(inhomogeneity). In the fully developed part of the flow, the Eulerian
statistics of the turbulence are stationary in time and
translation-invariant in the streamwise and transverse direction.
Thanks to these symmetries, the statistics can be conditioned on a
single parameter, the wall distance $y$.
This relative simplicity makes this flow a privileged framework to develop and
benchmark advanced Lagrangian stochastic models applicable to realistic flows.

\begin{figure}[!htb]
  \centering
  \includegraphics{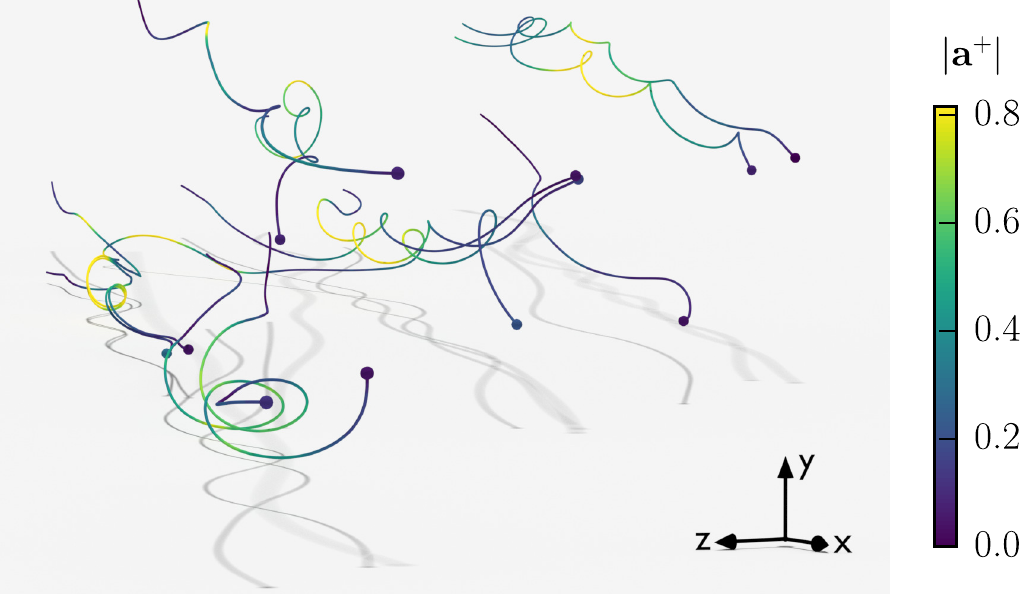}
  \caption{%
    Sample high-acceleration particle tracks obtained from DNS.\@
    Particles are located in the near-wall region ($y^+ \lesssim 200$).
    Trajectories are shown over $\Delta t^+ = 120$.
    The shadow is projected on the wall.
    Colors represent the norm of particle acceleration.
    $\vert \mathbf{a}^+ \vert=1$ corresponds roughly to \SI{430}{m/s^2} in the
    experiments.
  }\label{fig:trajectories}
\end{figure}

It is well known that near-wall turbulence is characterized by multiscale
coherent structures with preferential orientations~\cite{smits_highreynolds_2011}.
These structures include intense vortices elongated in the mean flow
direction, that strongly affect the near-wall flow dynamics.
These streamwise vortices induce strong centripetal accelerations, being the
main source of acceleration intermittency near the
walls~\cite{lee_intermittent_2004}, as illustrated on
Fig.~\ref{fig:trajectories}.
On the other hand, large-scale inhomogeneity implies that velocity and
acceleration statistics depend on wall distance.
Thus, it is also of interest to investigate the far-wall behavior, where a
return to isotropy may be expected, and stochastic models based on isotropic
turbulence may be applied.

We first present the experimental and numerical setups that allow us to measure the acceleration
of particles along their trajectories. In part~\ref{sec:correlations} we show the statistical analysis of the temporal
dynamics through the computation of time correlation functions of both acceleration and velocity 
components. This analysis provides estimates of the relevant time scales that are discussed in 
part~\ref{sec:timescales}. In part~\ref{sec:distributions} we focus on the acceleration probability distributions and compare them to the 
case of HIT. 

\section{Experimental and numerical setups}\label{sec:exp_methods}

\begin{figure*}[htb]
  \centering
  \includegraphics[]{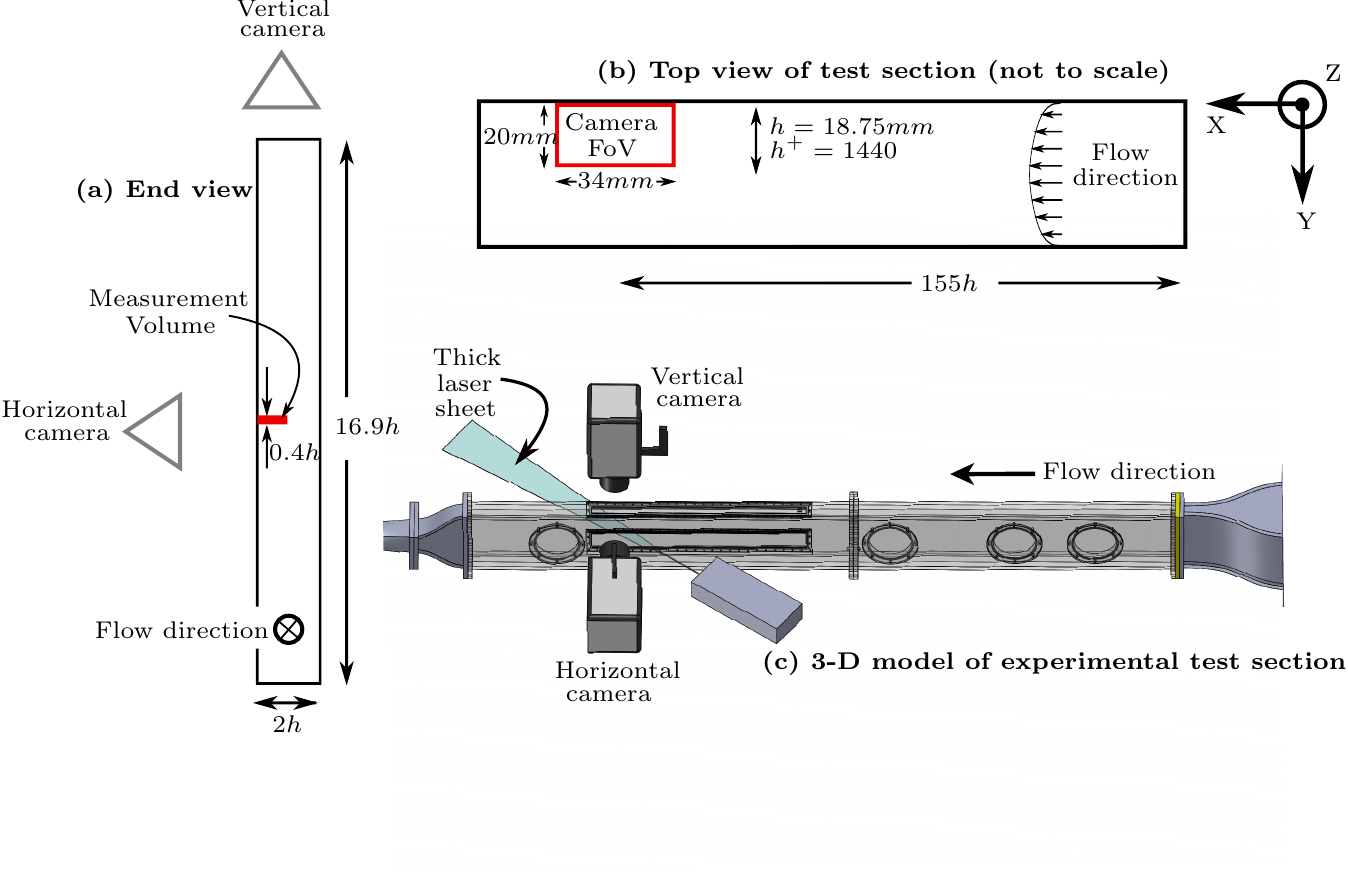}
  \vspace{-1.3cm}
  \caption{\label{fig:manip}Sketch of the turbulent channel used in the experiment. 
    Subfigure (a) is a sketch of the end view of the channel showing the aspect ratio 
    of the channel (the spanwise distance is $16.9h$) and the end view of the measurement 
    volume, as well as the position of the two cameras. Subfigure (b) shows a top view of the 
    channel, with the field of view (FoV) of the vertical camera highlighted. Subfigure (c) 
    shows a 3-D rendering of the experimental setup, including the relative positions of the 
    two high speed cameras and the thick laser sheet used to illuminate the measurement volume.
  }
\end{figure*}

We study the turbulent flow in a channel between two parallel walls
separated by a distance $2h$ using the same
Reynolds number
($\Rey = U_0 h / \nu = \num{34000}$) in both experiments and DNS\@.
This corresponds to a friction Reynolds number
$\Retau = u_\tau h / \nu \approx 1440$, where $u_\tau=\sqrt{\tau_w/\rho}$ is the
friction velocity associated to the shear stress $\tau_w$ at the wall and $\nu$ the
kinematic viscosity.
In the following, the superscript $+$ indicates quantities expressed in wall
units, nondimensionalized by $u_\tau$ and $\nu$.

The experiment consists of measurements made in a closed-loop water tunnel,
 shown in Fig.~\ref{fig:manip}, with a centerline velocity $U_0=\SI{1.75}{\meter\per\second}$.
 We chose water as a
working fluid in order to have neutrally buoyant and small enough tracer
particles, which is very difficult to achieve in air. The experimental test
section is \SI{3.2}{\m} long with a cross-section of
\SI{37.5 x 316}{\mm}, with tripped boundary layers at the entrance.
The development length is $155h$ and the channel height is $16.9h$, ensuring
statistical homogeneity in the streamwise and spanwise directions.

The wall unit is $\delta = \nu/u_\tau = \SI{13}{\micro\metre}$ in our
experimental conditions, thus we chose to seed the flow with
\SI{10}{\micro\metre} polystyrene spheres that are small enough to
accurately trace the flow down to the viscous layer.
The Stokes number of these particles ranges from $St = 0.02$ at $y^+ = 0.5$ to
$St=9 \times 10^{-4}$ at the center of the channel. Fluorescent
particles are used in order to improve the contrast in the vicinity of the wall by
eliminating reflections of the illumination laser near the wall. This choice
makes the measurement conditions quite challenging due to the weak amount of
light emitted by the particles.
Three dimensional particle trajectories are measured by particle tracking
velocimetry~\cite{ouellette} in a \SI{35 x 20 x 8}{mm} measurement volume illuminated by a
\SI{8}{mm}-thick 25W CW laser sheet, using two highly sensitive very high speed
Phantom v2511 cameras running at a sampling
rate of \SI{25000}{frames/s} (with one \SI{180}{mm} and one \SI{150}{mm} macro lenses with optical filters 
tuned to the emission frequency of the fluorescent particles). The measurement volume
covers half the width of the channel and is long enough in the downstream
direction for a sufficiently long time of particle tracking.
This allows us to observe the full decorrelation of the acceleration and (close to the wall) the
velocity.
Such a high sampling
rate is required in order to have enough time resolution to differentiate twice
the trajectories and compute the acceleration. Particle velocity and acceleration are
obtained by convolution of the
trajectories with Gaussian differentiating kernels, which also serves to filter
out noise from the measurements~\cite{mordant_experimental_2004}. The pixel size
corresponds to \SI{27}{\micro\metre} in physical space, but thanks to the diffraction of 
their emitted light, the fluorescent particles cover about three pixels in the images, which 
has been shown to be a good condition for subpixel position accuracy.
Indeed, the estimated accuracy is 1/10th
of a pixel (i.e.\ \SI{3}{\micro\metre}) after filtering. Although this allows us to have
a fairly precise estimation of the position of the particles in the bulk of the
flow, the apparent size of the particle and the existence of images reflected in
the wall prevents us from measuring the position of the particles very near the
wall.
The closest distance at which accurate detection of the
particle was possible is $y^+=4$, i.e.\ about \SI{50}{\micro\metre}. Thus,
our range of measurement spans the interval $y^+\in[4,1400]$ i.e.\ more than two
orders of magnitude in wall distance.

Direct numerical simulations are performed using a pseudo-spectral method for
the resolution of the velocity field between two parallel walls, coupled with
Lagrangian tracking of passive tracers advected by the resolved fluid velocity.
The pseudo-spectral method, described in detail by
\textcite{buffat_efficient_2011}, assumes periodicity in the streamwise ($x$)
and spanwise ($z$) directions, where a Fourier decomposition of the velocity
field is applied.
In the wall-normal ($y$) direction, a Chebyshev expansion is performed in order
to enforce no-slip boundary conditions at the walls.
The size of the computational domain is
$L_x \times L_y \times L_z = 4\pi h \times 2h \times \pi h$
(in wall units, $L_x^+ \times L_y^+ \times L_z^+ = 18166 \times 2891 \times 4541$)
in the streamwise, wall-normal and spanwise directions, respectively.
The velocity field is decomposed into $2048 \times 433 \times 1024$ spectral
modes.
In physical space, this corresponds to a uniform grid spacing $\Delta x^+ = 8.9$ and
$\Delta z^+ = 4.4$ in the streamwise and spanwise directions, respectively.
In the wall-normal direction, the grid spacing $\Delta y^+$ varies between
$0.04$ (wall region) and $10.5$ (channel center).
An explicit second-order Adams-Bashforth scheme is used to advance the resolved
equations in time, with a simulation time step $\Delta t^+ = 0.03$.
The total simulation time in channel units is $T U / h = 217$, which corresponds
to about $17$ turnover times of the centerline flow.

Once the instantaneous velocity field $\vec{u}$ is known, the acceleration field
is computed in the Eulerian frame according to
$\vec{a} = \partial\vec{u}/\partial t + \nabla \left( \vec{u}^2 / 2 \right) +
(\nabla \times \vec{u}) \times \vec{u}$.
Orzag's 2/3 rule~\cite{orszag_elimination_1971} is applied in the $x$ and $z$
directions to the velocity and acceleration fields to filter out
aliasing noise resulting from evaluation of non-linear terms.

The simulation is started with a fully-developed, statistically stationary
turbulent channel flow containing $2 \times 10^6$ randomly distributed fluid
particles.
Velocity and acceleration of fluid particles are determined from interpolation
of the respective Eulerian fields at each particle location using third-order
Hermite polynomials.
The choice of the interpolation scheme is critical, particularly for the evaluation
of Lagrangian acceleration statistics.
Lower-order schemes such as trilinear or Lagrange interpolation lead to
spurious oscillations which are clearly visible in the temporal spectrum of
particle
acceleration~\cite{choi_lagrangian_2004, van_hinsberg_optimal_2013}.
Particle positions are advanced in time using a second-order Adams-Bashforth
scheme, as for the Eulerian velocity field.
Sample trajectories obtained from this procedure are shown in
Figs.~\ref{fig:trajectories} and~\ref{fig:lagstat}.

\begin{figure}[!htb]
  \centering
  \includegraphics{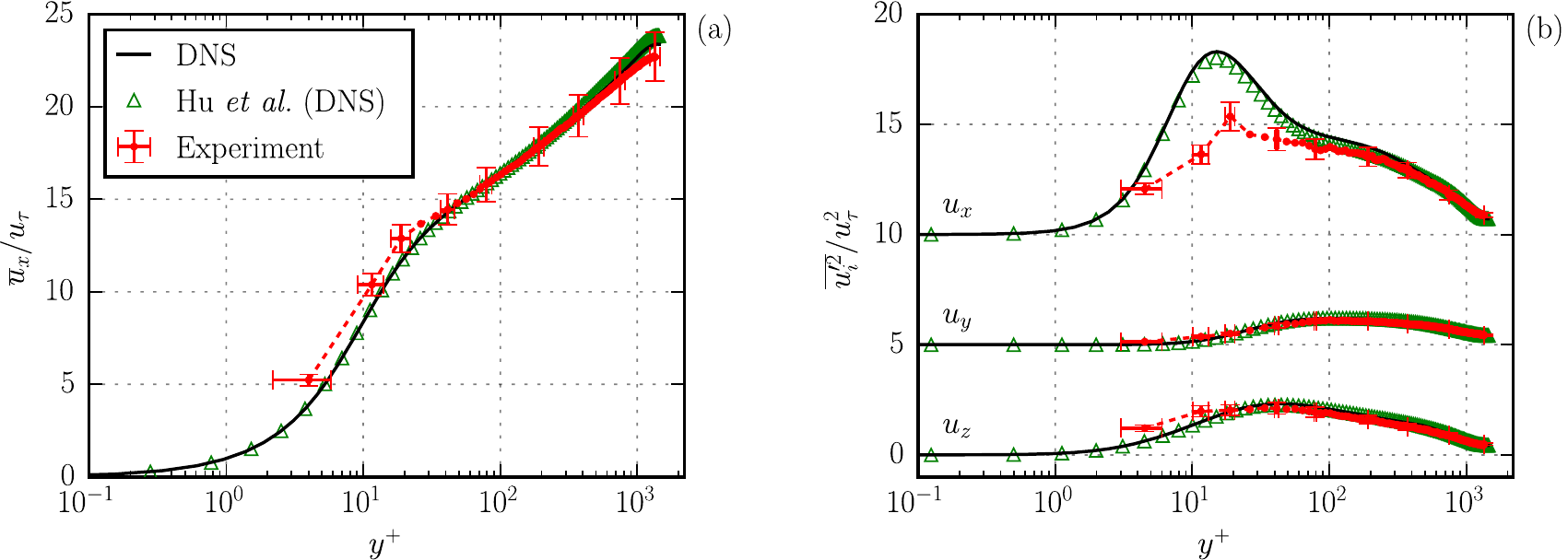}
  \caption{\label{fig:profiles_vel}%
    Mean and variance velocity profiles.
    Comparison between experiments (dashed lines), DNS (solid lines) and
    \textcite{hu_wall_2006} DNS at $\Retau = 1440$ (triangles).
    Velocity variance profiles are shifted vertically for clarity.
    All quantities are normalized in wall units.
  }
\end{figure}

In Fig.~\ref{fig:profiles_vel}, mean and variance velocity profiles from
experiments and simulations are compared with the channel flow DNS of
\textcite{hu_wall_2006} at roughly the same Reynolds number $\Retau = 1440$.
Experimental profiles are obtained by sampling the instantaneous velocity of
particles conditioned by their wall distance $y^+$.
In the three cases, the mean streamwise velocity profile presents a clear
logarithmic behavior over $40 \lesssim y^+ \lesssim 1200$.
Results from both simulations are consistent with each other, while slight
departures in the mean profile are observed for the experiments.
These differences are more pronounced near the wall ($y^+ < 30$) and towards the
channel center ($y^+ > 500$).
Similar remarks can be made for the streamwise velocity variance, where an
important difference is found at $y^+ < 50$ relative to the simulations.
On the other hand, wall-normal and transverse velocity variances from
experiments are in agreement with the simulations at all measured wall
distances.

\begin{figure}
  \centering
  \includegraphics{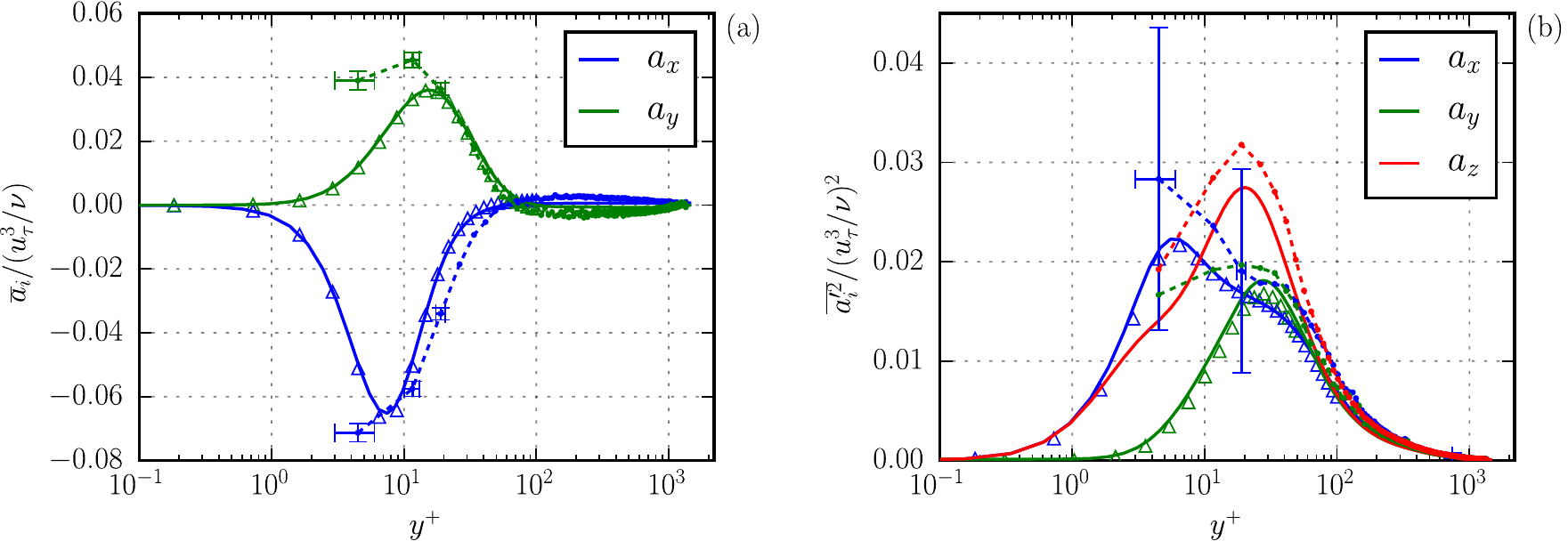}
  \caption{\label{fig:profiles_acc}%
    Mean and variance acceleration profiles.
    Comparison between experiments (dashed lines), DNS (solid lines) and
    \textcite{yeo_near-wall_2010} DNS at $\Retau = 600$ (triangles).
  }
\end{figure}

Error bars shown on the experimental results in Fig.~\ref{fig:profiles_vel} and the 
following experimental results reported in this paper (with the exception of the 
probability density function results discussed in Section \ref{sec:distributions}) are 
calculated statistically for a 95\% confidence interval~\cite{benedict_towards_1996}. 
Error bars also take into account the experimental precision associated with the parameters
used to report normalized results, $u_\tau$, $\nu$, etc., which is incorporated into the error
calculation in the standard way~\cite{moffat_robert_j._describing_1988}, and in some cases is responsible
for a large part of the error, as in the plot of $\overline{a'^2}/(u_\tau^3/\nu)^2$ shown in Fig.~\ref{fig:profiles_acc}. 

Figure~\ref{fig:profiles_acc} shows the mean and variance acceleration profiles
obtained by our experiments and DNS\@.
The profiles are consistent with the DNS results of
\textcite{yeo_near-wall_2010} (also presented in the figure) even though their
simulations were performed at a considerably lower Reynolds number
$\Retau = 600$.
As shown by \textcite{yeo_near-wall_2010}, the mean streamwise acceleration can
be decomposed into an irrotational and a solenoidal contribution, associated
with the mean streamwise pressure gradient and the viscous stress, respectively.
In wall units, this is expressed as
$\overline{a}_x^+ = \overline{a}_x^{I+} + \overline{a}_x^{S+} =
\frac{1}{\Rey_\tau} + \frac{d^2 \overline{u}_x^+}{d y^{+2}}$.
Near the wall, the solenoidal term $\overline{a}_x^{S}$ dominates and is
negative, which shows that the negative peak of mean streamwise acceleration at
$y^+ \approx 7$ is a consequence of a viscous contribution.
For the mean wall-normal acceleration, the solenoidal term
$\overline{a}_y^S$ is zero.
Therefore, its profile is entirely determined by the mean wall-normal pressure
gradient~\cite{yeo_near-wall_2010}.

Profiles of acceleration variance (Fig.~\ref{fig:profiles_acc}b) reveal
qualitative agreement between both sets of data, although large uncertainty is seen in the experimental results near the wall.
It is worth noting that, at their respective peaks, the standard deviation of acceleration
is larger than the magnitude of the mean acceleration, indicating that dynamics
near the wall are strongly influenced by acceleration fluctuations.
As shown by \textcite{lee_intermittent_2004}, these dynamics are dominated by
the presence of near-wall streamwise vortices inducing high-magnitude,
oscillating centripetal accelerations mainly oriented in the spanwise and
wall-normal directions.

\section{Lagrangian correlations}
\label{sec:correlations}

\begin{figure}[!htb]
  \centering
  \includegraphics{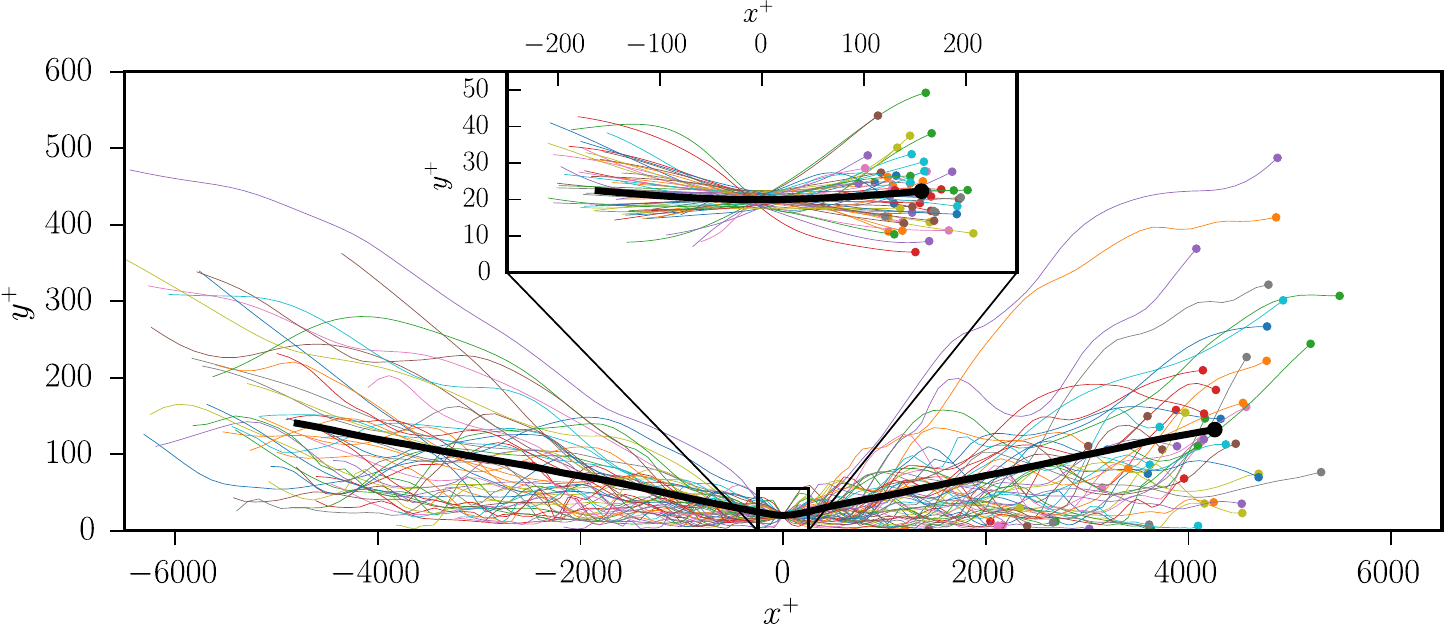}
  \caption{\label{fig:lagstat}%
    Illustration of the Lagrangian averaging procedure.
    Thin curves represent trajectories of particles located at
    $y^+ = y^+_0 \pm 0.5 \, \delta y^+$ at a reference time $t_0$ (here,
    $y^+_0 = 20$ and $\delta y^+ = 5$).
    Trajectories are shifted in the streamwise direction so that $x(t_0) = 0$.
    The thick curve represents the Lagrangian average of particle position
    $\langle \vec{r}(\tau, y_0) \rangle$.
    The channel center is at $y^+ = 1440$.
    Trajectories $\vec{r}(t_0+\tau, y_0)$ are shown for time lags
    $\tau^+ \in [-338, 338]$.
    The zoomed-up inset represents time lags $\tau^+ \in [-13.5, 13.5]$.
  }
\end{figure}

The Lagrangian description deals with particle trajectories that are
parameterized by their initial position $\vec{r}_0$ and by the time delay $\tau$
relative to the initial time $t_0$.
In stationary HIT, Lagrangian statistics do not depend on $\vec{r}_0$ due to
translational invariance, nor on $t_0$ due to statistical stationarity.
Statistics are thus parametrized only by the time delay $\tau$.
Furthermore, single-point single-time statistics such as moments of acceleration
components $\langle a_i^p(\tau) \rangle$ are constant in $\tau$.
In inhomogeneous turbulence, things are more complex.
Indeed the initial position $\vec{r}_0$ must be retained.
The symmetries of the channel flow are such that the dependency of the
statistics on $\vec{r}_0$ reduces to a dependency on the initial distance from
the wall, $y_0$.
Moreover, single-time Lagrangian statistics now vary with the time delay $\tau$.
For instance, the Lagrangian average of the streamwise velocity component,
$\langle v_x(t_0+\tau,y_0) \rangle$, depends on $\tau$ because particles move
away from their initial distance (toward the center on average, see
Fig.~\ref{fig:lagstat}) and thus experience regions of higher average velocity.

In inhomogeneous flows, the Lagrangian and Eulerian averages coincide only for
$\tau = 0$.
Stationarity of Eulerian statistics implies that Lagrangian statistics depend
only on $\tau$ and $y_0$ and not on $t_0$.
Inhomogeneity also implies that Lagrangian statistics for negative values of
$\tau$ are \textit{a priori} different from those at positive $\tau$.
Practically, estimators of Lagrangian statistics are the following.
A small interval of width $\delta y$ around a given initial value of $y_0$ is
chosen.
As soon as a trajectory has a value $y(t)$ that belongs to this interval,
the initial time $t_0$ is set.
Statistics are then accumulated as a function of $\tau$.
This procedure is illustrated in Fig.~\ref{fig:lagstat} for the average particle
position $\langle \vec{r}(\tau, y_0) \rangle$.

\begin{figure*}[!htb]
  \centering
  \includegraphics{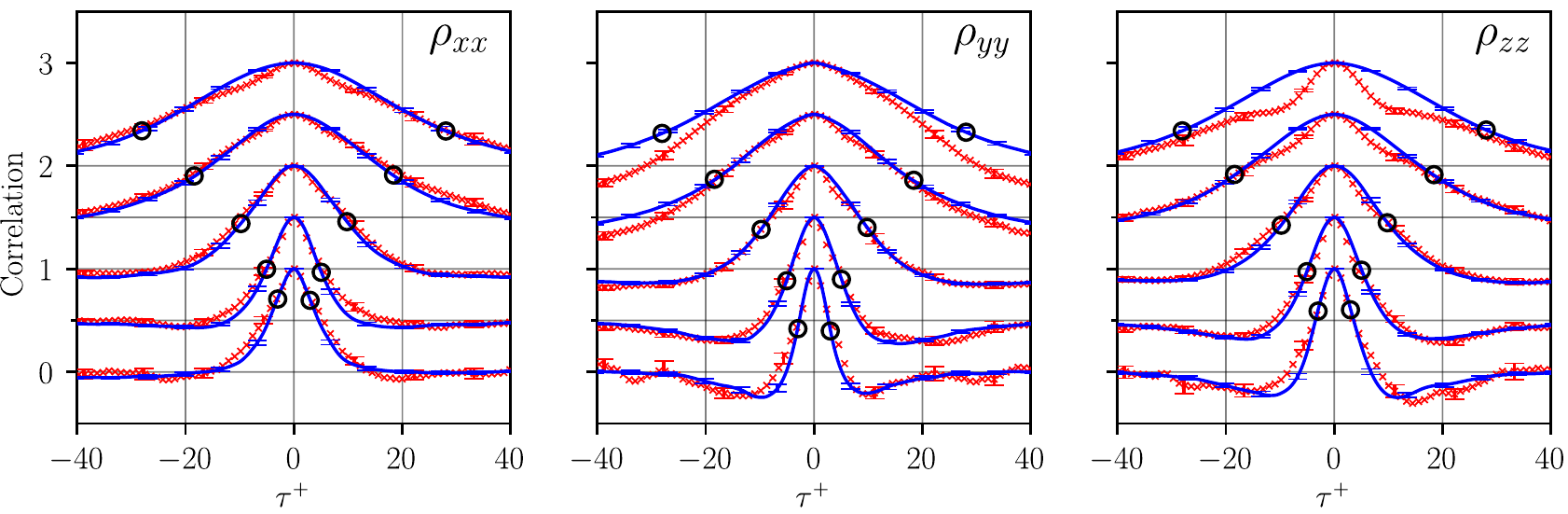}
  \caption{\label{fig:autocorrelations}%
    From left to right, Lagrangian auto-correlations of streamwise ($\rho_{xx}$),
    wall-normal ($\rho_{yy}$) and spanwise ($\rho_{zz}$) particle acceleration.
    Experiments - crosses.
    DNS - lines.
    Circles indicate time lags $\tau = \pm \tau_\eta$.
    Curves are shifted vertically by increments of 0.5 for clarity.
    From bottom to top, the curves correspond to particles located initially
    at $\yplus_0 = $ 20, 60, 200, 600 and 1000.
    Horizontal grid lines
show the zero-correlation level for each $\yplus_0$.
    In the experiment, $\tau^+ = 1$ corresponds to \SI{0.175}{milliseconds}.
  }
\end{figure*}

\begin{figure*}[t]
  \centering
  \includegraphics{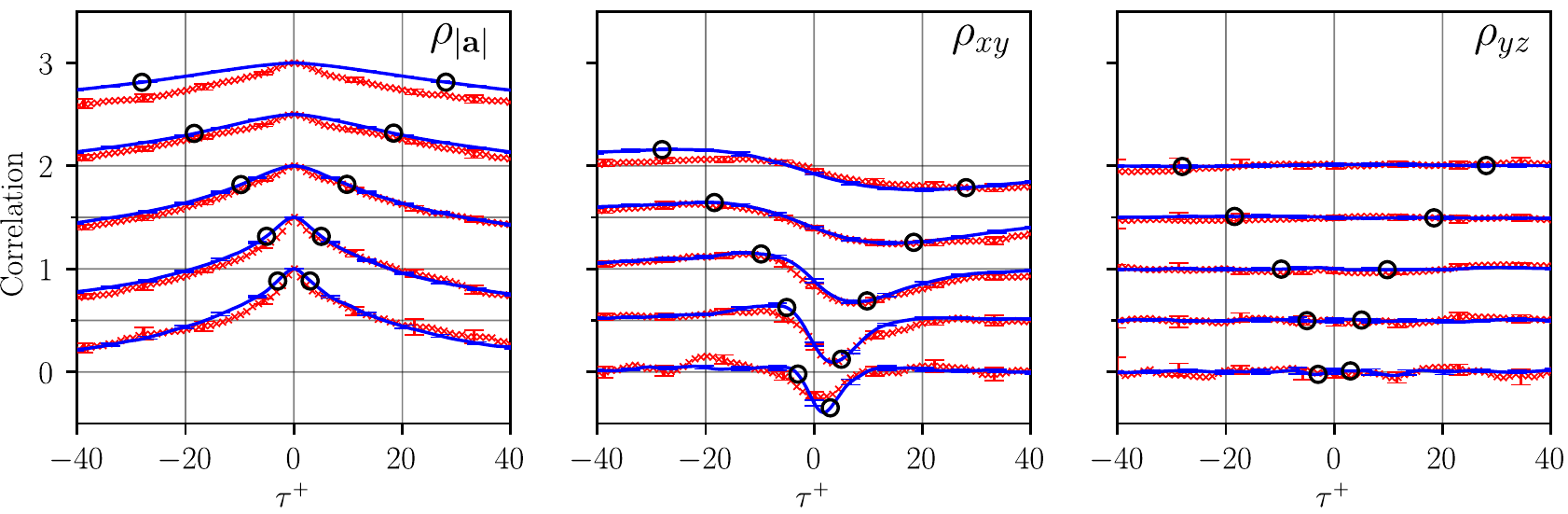}
  \caption{\label{fig:crosscorrelations}%
    From left to right, Lagrangian auto-correlation of acceleration magnitude
    ($\rho_{\anorm}$) and acceleration cross-correlations
    $\rho_{xy}$ and $\rho_{yz}$.
    From bottom to top, the curves correspond to particles located initially
    at $\yplus_0 = $ 20, 60, 200, 600 and 1000.
    (For details, see Fig.~\ref{fig:autocorrelations}.)
  }
\end{figure*}

An adequate tool to obtain time scales and coupling between components is the
Lagrangian correlation coefficient of fluid particle acceleration, defined as
\begin{equation}
  \rho_{ij}(\tau, y_0) =
  \frac
  {\left< a_i'(t_0,y_0) \, a_j'(t_0 + \tau,y_0) \right>}
  {{\left< a_{i\vphantom{j}}'^2(t_0,y_0) \right>}^{1/2} \,
  {\left< a_j'^2(t_0 + \tau,y_0) \right>}^{1/2}}\, \,,
\end{equation}
where $a_i'(t_0+\tau,y_0) = a_i(t_0+\tau,y_0) - \mean{a_i(t_0+\tau,y_0)}$ is the
fluid particle acceleration fluctuation relative to the Lagrangian average,
with $i = x, y$ or $z$.
The estimators thus correlate the initial acceleration with that at a time lag
$\tau$ along the trajectory of fluid particles initially at $y_0$.

\begin{figure}
  \centering
  \includegraphics{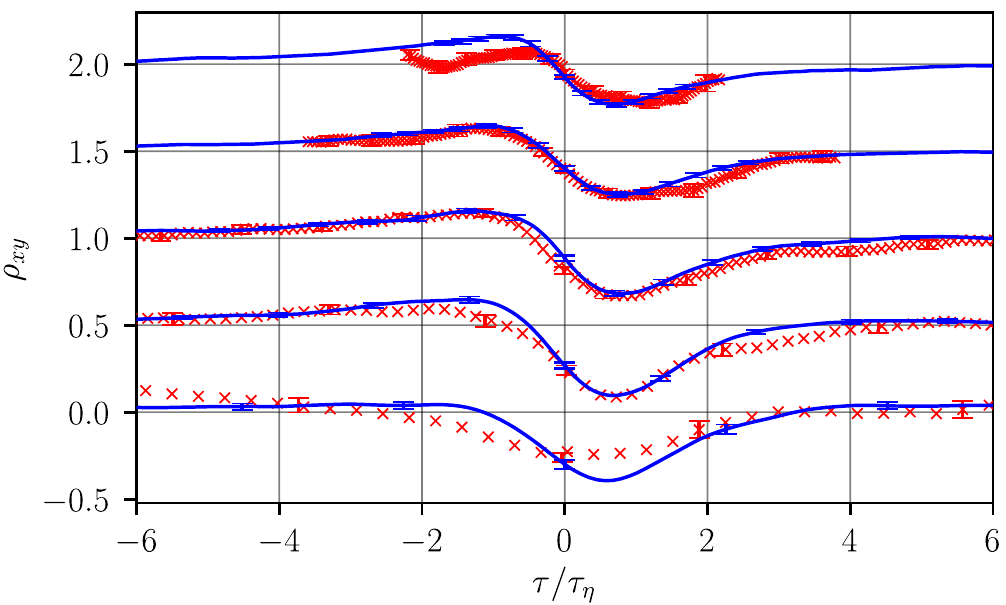}
  \caption{\label{fig:rho_xy_kolmog}%
    Lagrangian correlation between streamwise and wall-normal acceleration
    components.
    Time delay $\tau$ is normalized with the local Kolmogorov time scale
    $\tau_\eta$.
    From bottom to top, the curves correspond to particles located initially
    at $\yplus_0 = $ 20, 60, 200, 600 and 1000.
    (For details, see Fig.~\ref{fig:autocorrelations}.)
  }
\end{figure}

Figures~\ref{fig:autocorrelations} and~\ref{fig:crosscorrelations} show various
components of the acceleration correlation tensor $\rho_{ij}$ calculated at
different initial wall distances $y_0^+$.
Time lags equal to the local Kolmogorov time scale
$\tau_\eta(y) = \sqrt{\nu/\epsilon(y)}$ are also represented in the figures,
with the local mean turbulent energy dissipation rate estimated from DNS as
$\epsilon = \nu \overline{(\partial_j u_i') (\partial_j u_i')}$.

The inhomogeneity of the flow is visible in
the fact that the typical decorrelation time varies significantly over the width
of the channel. The anisotropy is visible in the fact that the streamwise and
wall-normal components display some non zero correlation (in contrast with HIT),
as shown on Fig~\ref{fig:crosscorrelations}(b). Cross-correlations with the transverse 
component remain zero due to the
statistical symmetry $z \leftrightarrow -z$, as shown on
Fig.~\ref{fig:crosscorrelations}(c).

In the vicinity of the wall the decorrelation time is close to one in wall
units, showing that this is the adequate characteristic time for rescaling of
small-scale quantities such as the acceleration in this region.
This will be discussed further in the analysis of the characteristic Lagrangian
time scales. A very good agreement is observed between experimental and DNS results, with the exception 
of the long time behavior of $\rho_{yy}$ at $y^+=1000$, and the short time behavior of 
$\rho_{zz}$ at $y^+=1000$. While the former remains unexplained, the latter is
due to the higher level of noise in the measurement of the $z$-component, which
is a technical consequence of the way the PTV is performed.
Namely, near the center of the channel, the signal-to-noise ratio for the $z$-component of acceleration comes close to the limits of the 
data processing methods described in Section~\ref{sec:exp_methods} and in
previous work~\cite{mordant_experimental_2004}, and the correlation of noise is
seen in the short time behavior of $\rho_{zz}$ at $y^+=1000$.

Previous studies in HIT~\cite{mordant_exp_num_2004, toschi_acceleration_2005} have
associated fluid particle acceleration and vortex dynamics by observing that
high acceleration events often correspond to centripetal accelerations in vortex
filaments, and the auto-correlation of the centripetal component of these
accelerations become negative to a much greater degree than the auto-correlation
of the component of the acceleration parallel to the vortex filament.
Contrary  to HIT where there are no preferential directions, in the near-wall
region of a wall-bounded turbulent flow the preferential orientation
of vortices in the streamwise direction is expected.
As shown on Fig.~\ref{fig:autocorrelations}, near the wall the
auto-correlations of $a_y$ and $a_z$ become negative at approximately
$\tau = 2\tau_\eta$ (similarly to the case of HIT), while the correlation of
$a_x$ remains positive with a significantly longer initial decorrelation time.
The negative $\rho_{yy}$ and $\rho_{zz}$ correlations can be associated with the effect of near-wall streamwise vortices. A fluid particle trapped in one such vortex experiences strong centripetal
accelerations towards the vortex rotation axis~\cite{lee_intermittent_2004}. This strong form of anisotropy of the acceleration is only observed near the
walls ($y^+ < 50$) and becomes negligible towards the channel center, as confirmed below by the acceleration time scales associated to these correlations.

The auto-correlations $\rho_{xx}$ and $\rho_{yy}$ are almost symmetric in time,
e.g.\ $\rho_{xx}(-\tau, y_0) \approx \rho_{xx}(\tau, y_0)$.
This symmetry can be explained by the nearly time-symmetric average trajectories
in the near-wall region as illustrated in Fig.~\ref{fig:lagstat},
i.e.\ $\mean{\vec{r}(\tau, y_0)} \approx \mean{\vec{r}(-\tau, y_0)}$.
In contrast, the cross-correlation $\rho_{xy}$ is strongly time-asymmetric close
to the wall.
This asymmetry, along with the non-zero time lag at which the peak is
observed, suggests the idea of causality between acceleration components.
That is, a streamwise acceleration fluctuation is followed on average by
an opposite-sign wall-normal acceleration fluctuation.
Towards the channel center, this effect persists and the correlation becomes
antisymmetric with time, $\rho_{xy}(-\tau) \approx -\rho_{xy}(\tau)$, due to the
decreasing influence of wall confinement.

The correlation between $a_x$ and $a_y$ (Fig.~\ref{fig:crosscorrelations}(b)) is
most important near the walls.
In that region, the zero-time cross-correlation (equivalent to an Eulerian
single-point single-time correlation) is negative due to increased viscous
effects combined with confinement by the wall (see the joint PDFs in
Section~\ref{sec:distributions} for more details).
Moreover, the cross-correlation peak is always found at a non-zero time lag.
Far from the wall, the correlation is close to zero at $\tau^+ = 0$, while it
increases in absolute value for non-zero time lags.
The influence of the boundary layer remains visible even in the bulk of the
channel where the correlation is still non-zero, indicating that small-scale
anisotropy is still present in that region.
In Fig.~\ref{fig:rho_xy_kolmog}, the $\rho_{xy}$ cross-correlation is displayed
as function of the normalized time lag $\tau / \tau_\eta$.
The time lag of the negative correlation peak is shown to scale with
$\tau_\eta$, with a value $\tau / \tau_\eta$ fluctuating between $0.5$ and $0.7$
with the wall distance $y_0^+$.

The $\rho_{xy}(\tau, y_0)$ correlation describes
the changes of orientation of the acceleration fluctuation vector
$\vec{a}'(t_0+\tau, y_0)$ projected on the $x$-$y$ plane.
Its behavior implies that there is a preferential direction of rotation of
$\vec{a}'$ along a particle trajectory.
Moreover, such changes of orientation happen over times of the order of the
Kolmogorov time scale.
Thus, this anisotropy is associated with the smallest scales of turbulence, and is
observed for all wall distances.
The preferential direction of rotation implied by the
cross-correlation is consistent with the direction of mean shear, represented
by an average vorticity $\overline{\omega}_z = -d\bar{u}/dy$ which is negative in
the lower half of the channel, where the presented statistics are obtained.
This result is consistent with evidence of small-scale anisotropy found in other
turbulent flows governed by large-scale anisotropy.
For instance, from DNS of homogeneous shear flow,
\textcite{pumir_persistent_1995} found signs of small-scale anisotropy which did
not decrease at increasing Reynolds number, in contradiction with Kolmogorov's
local isotropy hypothesis~\cite{kolmogorov_local_1941}.
In their work, small-scale anisotropy was quantified by the skewness of the
spanwise vorticity $\omega_z$, which was shown to be of the same sign as the
large-scale average vorticity.
More recently, and using a similar approach, \textcite{pumir_small-scale_2016}
showed the presence of small-scale anisotropy from DNS of turbulent
channel flow at $\Retau \approx 1000$ all along the log-layer.
Our results show that such small-scale anisotropy is also observed by the
Lagrangian acceleration statistics.
Therefore, a stochastic model for the Lagrangian acceleration which includes
elements derived from the $\rho_{xy}$ correlation would be able to reproduce the
presence of small-scale anisotropy in shear flows.

In Fig.~\ref{fig:crosscorrelations}(a) we plot the auto-correlation of the
acceleration magnitude $\anorm$, showing that this quantity stays
correlated for far longer than each acceleration component.
This behavior is observed at all wall distances and is consistent with results
in HIT~\cite{yeung_lagrangian_1989, mordant_three-dimensional_2004}.
It is explained by the fact that changes in the orientation of the acceleration
vector are much more sudden than changes in its magnitude.
In near-wall turbulence, this observation is again explained by centripetal
acceleration induced by streamwise vortices, which preserve the acceleration
magnitude for a longer time than the acceleration
orientation~\cite{lee_intermittent_2004}.

\begin{figure}[t]   
	\centering
    \includegraphics[width=0.5\textwidth]{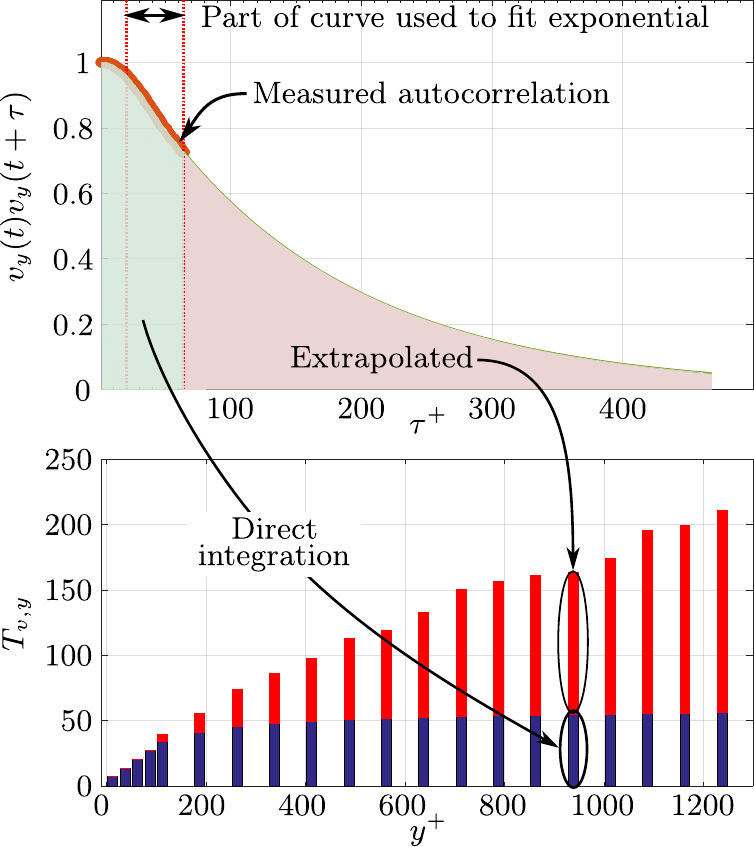}
	\caption{\label{fig:timescale_explain}%
		An example of how Lagrangian velocity
		time scales are calculated.   The auto-correlation
		of the wall-normal component of velocity from $y^+=925$ is shown    (top plot)
		in red, where the part of the measured auto-correlation used to fit the
		exponential    extrapolation is also shown. The bottom plot shows the Lagrangian
		wall-normal velocity time    scale across the channel, where the portion of the
		time scale that is directly measured    is shown in blue and the portion of the
		time scale that comes from the extrapolation   is shown in red.   } 
\end{figure}

\begin{figure}[t]
  \centering
  \includegraphics{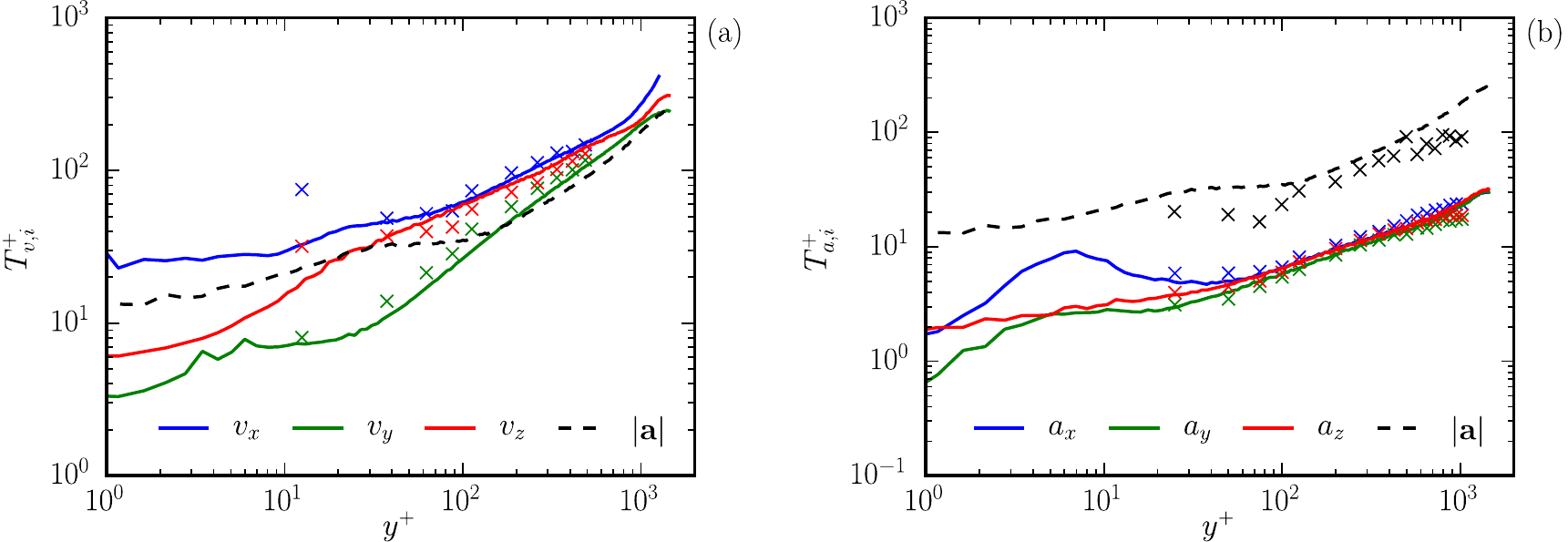}
  \caption{\label{fig:timescales_plus}%
    Lagrangian velocity and acceleration time scales in wall units.
    Experiments - crosses.
    DNS - solid lines.
    The acceleration magnitude time scale $T_{\anorm}$ is represented by dashed
    lines.
  }
\end{figure}

\begin{figure}
  \includegraphics{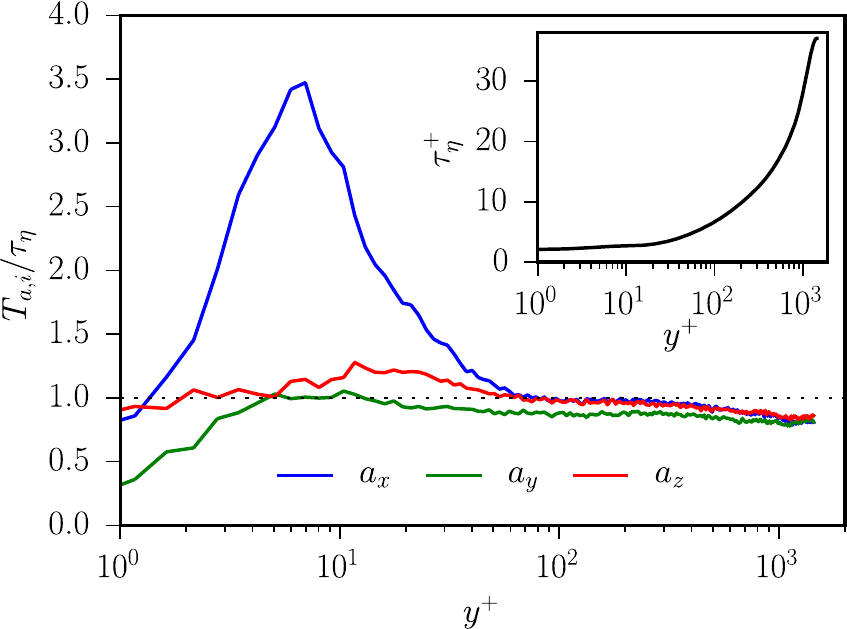}
  \caption{\label{fig:timescales_taueta}%
    Lagrangian acceleration time scales normalized by the local Kolmogorov time
    scale (DNS results only).
    Inset: local Kolmogorov time scale in wall units. 
  }
\end{figure}

\begin{figure}
  \includegraphics{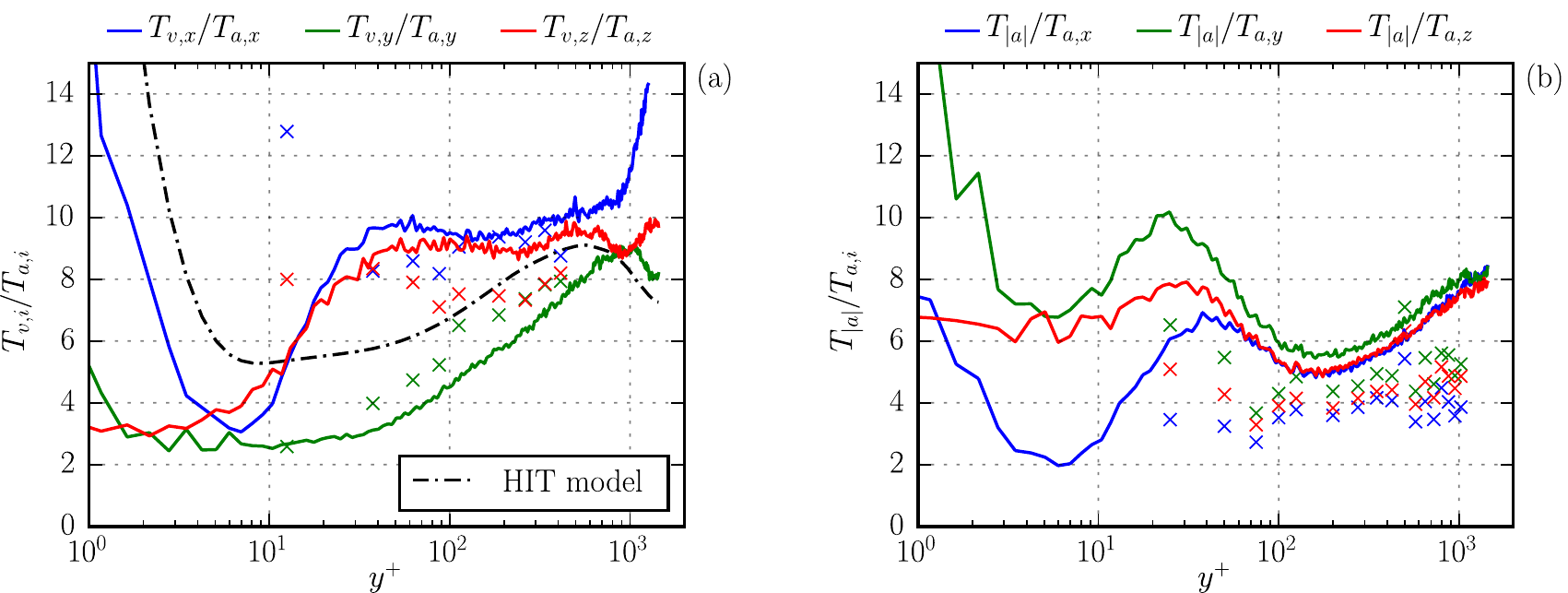}
  \caption{\label{fig:timescale_ratios}%
    Lagrangian time scale ratios. Experiments - crosses.
    DNS - solid lines.
    (a) Ratio between the Lagrangian velocity and acceleration time scales, by component.
    Also shown is the HIT model~\cite{sawford_reynolds_1991} for the ratio of Lagrangian velocity
    and acceleration time scales.
    (b) Ratio between the time scales of acceleration magnitude and the components of 
    acceleration.
  }
\end{figure}

\section{Lagrangian time scales}
\label{sec:timescales}
The characteristic time scale associated to each acceleration component is
estimated according to
\begin{equation}
  T_{a,i}(y_0) = \int_0^{\tau_c} \rho_{ii} (\tau, y_0) \, \mathrm{d} \tau,
  \label{eq:timescale_acc}
\end{equation}
where $\tau_c$ is the time lag at which the auto-correlation first crosses
\num{0.05}.
This definition is chosen because the classical definition with
the integration going to infinite time cannot be applied to all acceleration
components since some correlations become negative.
This is also the case for HIT for which the integral of the acceleration
correlations is actually zero because of the stationarity of the velocity.
The usual zero-crossing time as in \textcite{yeung_lagrangian_1989} can neither
be used here because some correlations (near the center of the channel) do not
cross zero during the observation time.
Our definition is a convenient mix between these two usual definitions of the
typical time scale.

Lagrangian velocity (integral) time scales $T_{v,i}$, as well as the acceleration
norm time scale $T_\anorm$, are defined equivalently. Due to the limited
measurement volume in the experiment, the full decorrelation of the auto-correlations of velocity 
and the norm of the acceleration is not achieved at all wall distances in the
channel. These auto-correlations have been extrapolated as illustrated by
Fig.~\ref{fig:timescale_explain}, and the uncertainty of these results increases with increased extrapolation.
This extrapolation is not necessary for the acceleration auto-correlations shown in
Fig.~\ref{fig:autocorrelations} (which decay much faster) and thus for the
computation of the time scales $T_{a,i}$.

The evolution of all time scales with wall distance is shown in
Fig.~\ref{fig:timescales_plus}. As can be deduced from the auto-correlation curves, the acceleration time scales
$T_{a,i}$ and $T_\anorm$ generally increase with wall distance.
The same is observed for the Lagrangian velocity time scales.
The acceleration norm time scale $T_\anorm$ is about one order of
magnitude larger than the time scale of the acceleration components.
It is of the order of the integral time scales $T_{v,i}$.

In Fig.~\ref{fig:timescales_taueta}, the Lagrangian acceleration time scales are
normalized with the local Kolmogorov time scale.
Both $\epsilon$ and $\tau_\eta$ vary with wall distance.
The acceleration time scales are of the order of $\tau_\eta$ all along the
channel.
The normalized time scales only weakly change for $y^+>80$, reaching a value
between 0.8 and 0.9 in the bulk of the channel.
However, in that region, small differences persist between $T_{a,y}$ and
the time scales obtained for the other two components, suggesting once more that
anisotropy is still present far from the wall.
Close to the wall ($y^+ \lesssim 40$), the longer correlation time of the
streamwise acceleration is reflected in a larger time scale $T_{a,x}$ compared to
the other components.

Fig.~\ref{fig:timescale_ratios}(a) shows the ratio of Lagrangian time scales of velocity and 
acceleration for each component.
Also shown in Fig.~\ref{fig:timescale_ratios}(a) is eq.~\eqref{eq:HIT_model}, 
the empirical fit to the DNS data of Yeung and Pope~\cite{yeung_lagrangian_1989} proposed
by Sawford~\cite{sawford_reynolds_1991}, using the profile of $Re_\lambda$
calculated from the DNS and $C_0 = 7$ as suggested by Sawford. 
The HIT model follows the trend of the data.
However there is significant anisotropy in these ratio of
time scales that extends far away from the wall.
First-order Lagrangian stochastic models in velocity such as given by
eq.~\eqref{eq:langevin} are based implicitly on the scale separation between the
velocity and the acceleration time scales.
Here, a small separation is seen between these two time scales near the wall.
Significantly, the time scale ratio for the wall-normal component ($y$) is
approximately half of that predicted by the local Reynolds number (the HIT model
plotted in the figure) near the wall.
Even farther from the wall, this time scale ratio is significantly
over-predicted by the HIT model.

The long time scales of the acceleration norm previously reported have inspired
the development of a Lagrangian subgrid stochastic model that models the
acceleration norm and acceleration direction as two independent
stochastic processes~\cite{sabelnikov_new_2011, zamansky_acceleration_2013}.
Fig.~\ref{fig:timescale_ratios}(b) shows that the ratio of Lagrangian time
scales of acceleration norm to acceleration components are only weakly varying
for $y^+>50$ and comparable in magnitude to the ratios of Lagrangian velocity
and acceleration time scales.

\section{Distributions}\label{sec:distributions}

\begin{figure*}[t]
  \centering
  \includegraphics{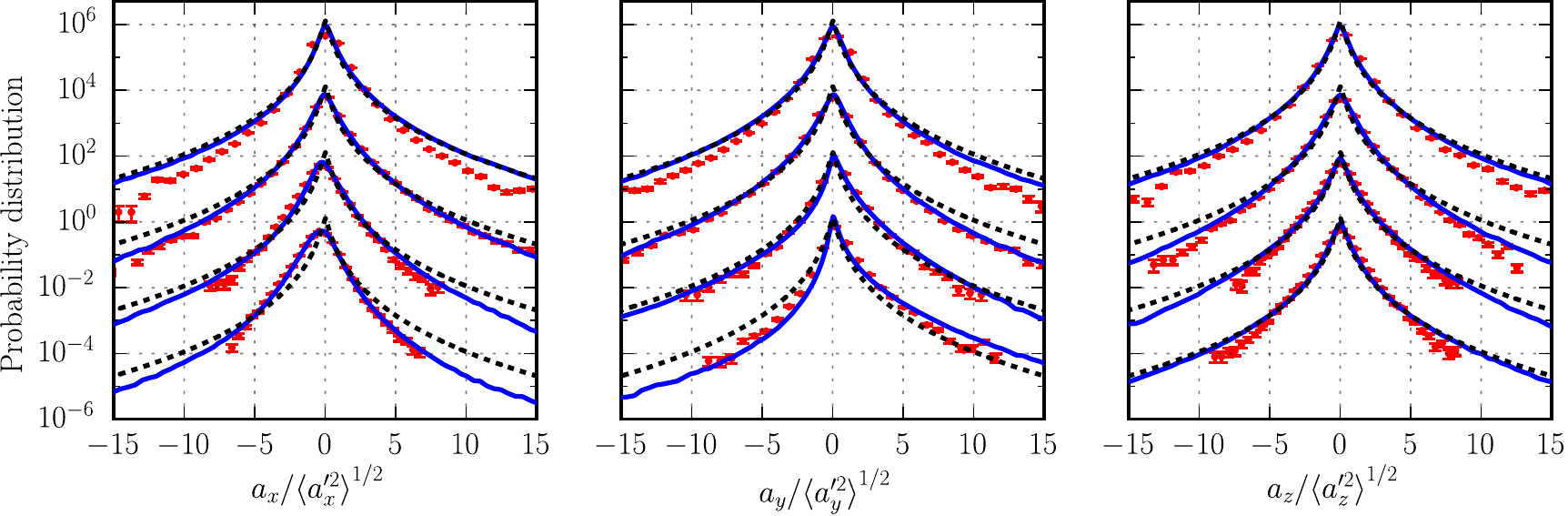}
  \caption{\label{fig:pdf_acc}%
    PDF of streamwise, wall-normal and spanwise particle acceleration.
    Experiments - symbols with error bars. DNS - lines.
    The dashed lines represent the theoretical prediction for the acceleration
    PDF in HIT~\cite{mordant_three-dimensional_2004}.
    The PDFs are normalized by the root-mean-square value of acceleration.
    From bottom to top, the curves correspond to particles located at
    $\yplus = 10$, $20$, $200$ and $1200$.
    The statistical convergence of the experimental data is shown by error bars
    proportional to $1/\sqrt{n_i}$ where $n_i$ is the number of events in bin $i$.
  }
\end{figure*}

\begin{figure}[htb]
  \centering
  \includegraphics{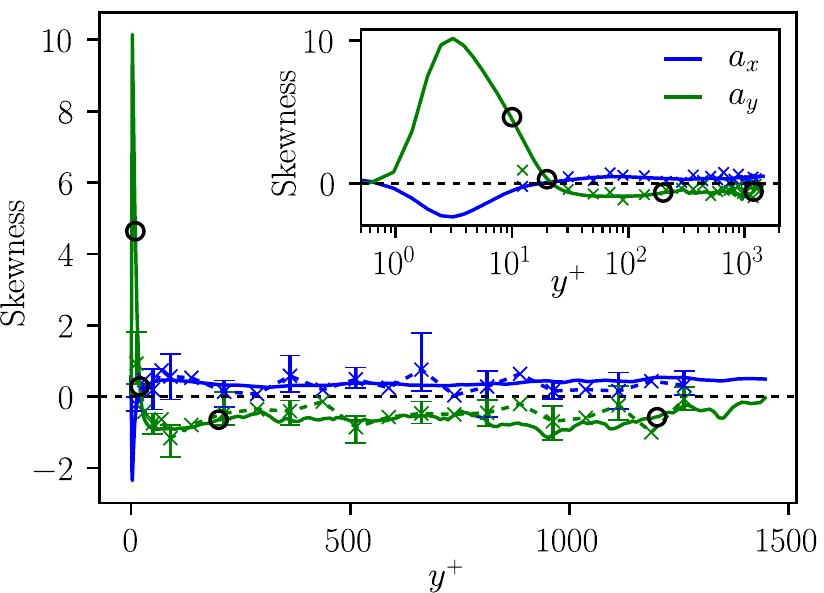}
  \caption{\label{fig:skewness_acc}%
    Skewness of streamwise and wall-normal acceleration components.
    Experiments - crosses. DNS - lines.
    Circles indicate skewness of $a_y$ at
    $\yplus = 10$, $20$, $200$ and $1200$.
    Inset: skewness profiles with $y^+$ in logarithmic scale.
  }
\end{figure}

Fig.~\ref{fig:pdf_acc} shows the probability distribution function (PDF)
of the three acceleration components obtained at different wall distances.
All curves present very long tails corresponding to extremely
high acceleration events associated to
intermittency~\cite{la_porta_fluid_2001}.
Once again, good agreement is achieved between the experiments and the DNS\@.
The acceleration PDFs are also compared with the functional shape proposed
by \textcite{mordant_three-dimensional_2004} for HIT, which assumes that the
acceleration magnitude follows a log-normal probability distribution and that
the acceleration vector is isotropic.
According to those assumptions, the PDF of an acceleration component is given by
\begin{equation}
  P(a_i) =
  \frac{e^{s^2/2}}{4m}
  \left[
    1 - \operatorname{erf}
    \left( \frac{\ln \frac{\lvert a_i \rvert}{m} + s^2}{\sqrt{2} s} \right)
  \right],
\end{equation}
where $m$ determines the variance of $a_i$ ($m = \sqrt{3/e^{2s^2}}$ for variance
1), while $s$ determines the shape of the PDF\@.
A value $s = 1$ is used in the comparisons.
Towards the channel center, the PDFs of the three acceleration components match
this prediction, suggesting that the instantaneous behavior of acceleration
becomes close to isotropic.
More strikingly, the spanwise acceleration seems to match the prediction very
close to the wall, suggesting that this component is not affected by anisotropy
as in the other two directions.
The general agreement with the shape of the HIT PDF suggests
that intermittency is extremely strong in the boundary layer although the Reynolds number 
is moderate: $Re_{\lambda}\sim 60$ to $100$ in our flow whereas it was close to 1000 in 
\textcite{mordant_three-dimensional_2004}. It implies also that this shape of the PDF 
presents some universality.

The PDFs of the streamwise and wall-normal acceleration components become quite
asymmetric near the wall.
This asymmetry is quantified by their skewness
$S_i = \mean{a_i'^3} / \mean{a_i'^2}^{3/2}$ shown in
Fig.~\ref{fig:skewness_acc}.
Due to flow symmetry, the skewness of $a_z$ is zero.
Very close to the wall, $S_x$ and $S_y$ are strongly negative and positive,
respectively, indicating that their PDFs are very asymmetric due to wall-induced
anisotropy.
The signs of $S_x$ and $S_y$ are both inverted after $y^+ \approx 20$.
Their respective values remain different from zero and change little for $y^+>80$.
This reinforces the idea that turbulence anisotropy is still present in the
channel center.

\begin{figure}[t]
  \centering
  \includegraphics{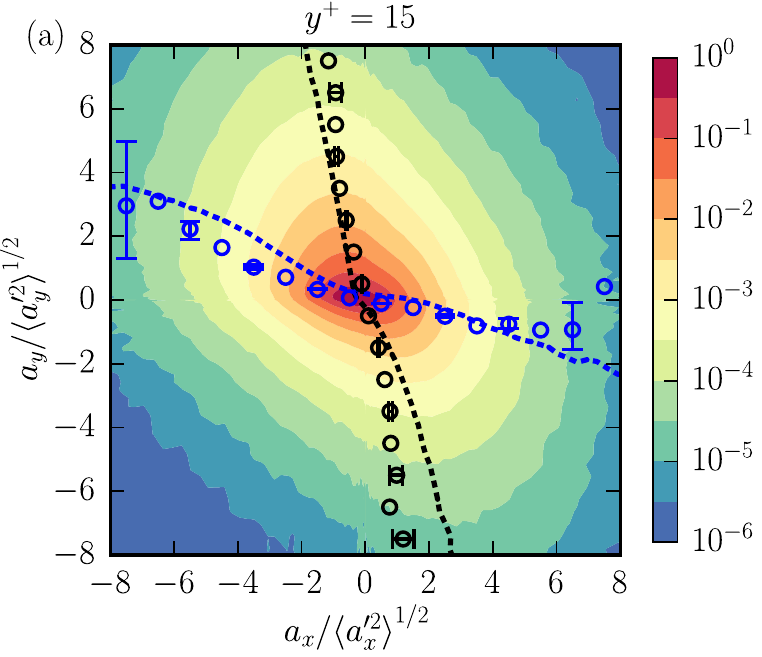}
  \hspace{1cm}
  \includegraphics{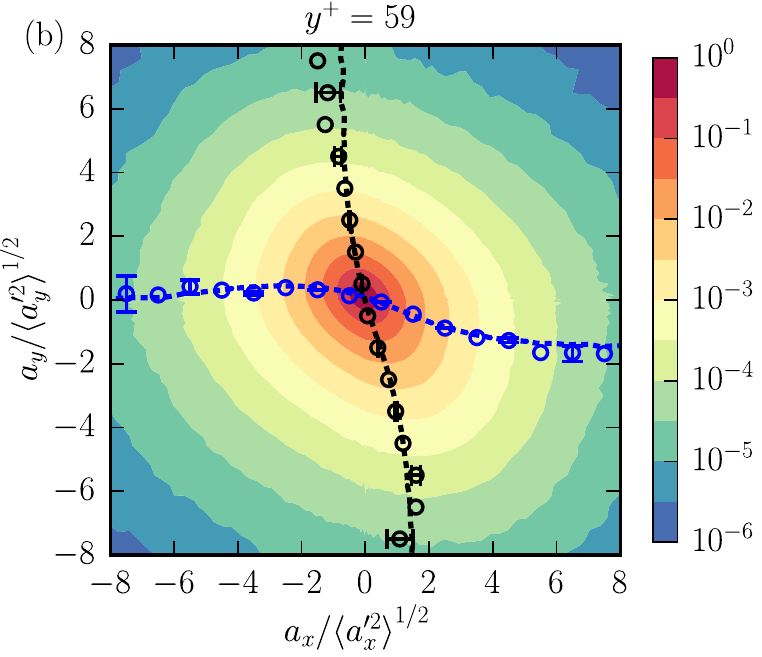}
  \caption{\label{fig:jointpdf}%
    Joint PDF of streamwise and wall-normal acceleration at
    $\yplus = 15$ and $59$.
    Conditional means $\mean{a_x | a_y} / \mean{a'^2_x}^{1/2}$ and
    $\mean{a_y | a_x} / \mean{a'^2_y}^{1/2}$ are superposed to the contours
    using black and blue markers, respectively.
    Experiments - circles.
    DNS - dashed lines.
  }
\end{figure}

The dependency between the acceleration components is analyzed by the joint PDF,
$P(a_x, a_y)$.
Fig.~\ref{fig:jointpdf} shows the results obtained at two wall distances,
$y^+ = 15$ and $59$.
Close to the wall (Fig.~\ref{fig:jointpdf}a), the joint PDF has a stretched
shape, showing a preference for events given by $a_x < 0$ and $a_y > 0$.
These events can be associated to fluid particles advected towards the wall.
These particles
move towards regions of decreasing mean streamwise velocity,
leading on average to a streamwise deceleration of their motion ($a_x < 0$).
Simultaneously, their negative wall-normal velocity goes to zero due to
confinement by the wall, resulting in a positive wall-normal acceleration
($a_y > 0$).
This dependency between both acceleration components is confirmed by the
conditional means $\mean{a_x | a_y}$ and $\mean{a_y | a_x}$, which are
superposed to the joint PDF contours.
Particles advected away from the wall are less affected by
wall confinement, and thus the impact on their wall-normal acceleration
is less visible in the joint PDF\@.
Nonetheless, on average, those particles also experience the effect of the mean
velocity gradient, which results in this case in a positive streamwise
acceleration.

\section{Concluding remarks}
\label{sec:conclusions}

By performing both DNS and experiments, the highly anisotropic turbulent flow in
the vicinity of channel walls is described in terms of Lagrangian statistics.
Near-wall vortical structures are clearly identified by their influence on the
acceleration auto-correlations.
Viscous effects and wall-confinement have also a strong
impact on acceleration statistics as evidenced by the joint PDFs.
Less expectedly, signs of small-scale anisotropy are present across the channel.
The observed behavior of the Lagrangian time scales can be a basis for the
formulation of Lagrangian stochastic models of acceleration applied to
wall-bounded turbulent flows.
These models must also take into account the dependency among acceleration
components, which is apparent in the presented cross-correlations and joint
PDFs of acceleration.
Acceleration models based on the presented results should be able to capture the
effects of (i) the near-wall dynamics associated with confinement and coherent
structures and (ii) the small-scale anisotropy present in the whole channel.

\begin{acknowledgments}
  This work is supported by Agence Nationale de la Recherche
  (grant \#ANR-13-BS09-0009) and by the LabEx Tec 21
  (Investissements d'Avenir grant \#ANR-11-LABX-0030).
  Simulations have been performed on the P2CHPD cluster.
  J.I.P.\ is grateful to CONICYT Becas Chile grant No.\ 72160511 for supporting
  his work.
  NM is supported by Institut Universitaire de France.
  We thank Laboratoire de Physique at ENS de Lyon and CNRS for providing one of
  the high speed cameras.
  We thank M. Kusulja for the design of the water tunnel, J. Virone and V.
  Govart for technical assistance.
\end{acknowledgments}

\bibliography{biblio.bib}

\end{document}